\documentclass{svjour2}                    
\smartqed  
\usepackage{graphicx}
\usepackage{amssymb}
\newcommand{\be}{\begin{equation}} 
\newcommand{\ee}{\end{equation}}
\newcommand{\beq}{\begin{eqnarray}}
\newcommand{\eeq}{\end{eqnarray}}
\newcommand{\bt}{\beta}
\newcommand{\bl}{\begin{lemma}}
\newcommand{\el}{\end{lemma}}
\newcommand{\bm}{\begin{pmatrix}}
\renewcommand{\em}{\end{pmatrix}}
\newcommand{\bml}{\begin{multline}}
\newcommand{\eml}{\end{multline}}
\newcommand{\ba}{\begin{array}}
\newcommand{\ea}{\end{array}}

\newcommand{\la}{\label}
\newcommand{\ci}{\cite}

\newcommand{\al}{\alpha}

\renewcommand{\th}{\theta}

\newcommand{\bi}{\bibitem}

\newfont{\msbm}{msbm10 scaled\magstep1}
\newfont{\msbms}{msbm7 scaled\magstep1} 

\newcommand{\bbc}{\mbox{$\mbox{\msbm C}$}}

\journalname{Journal of Statistical Physics}
\begin{document}

\title{The Fisher-Hartwig Formula and Generalized  Entropies  in XY Spin Chain
}


\author{A. R. Its         \and
        V. E. Korepin 
}


\institute{A. R. Its \at
              Indiana University - Purdue University Indianapolis \\
              \email{itsa@math.iupui.edu}           
           \and
           V. E. Korepin \at
              C.N. Yang Institute for Theoretical Physics\\
 							State University of New York at Stony Brook,
 							Stony Brook, NY 11794-3840, USA\\
 							Tel.: +1-631-632-7981\\
              Fax: +1-631-632-7954\\
              \email{korepin@gmail.com}
}

\date{Received: date / Accepted: date}

\maketitle

\begin{abstract}
Toeplitz matrices have applications to different problems of statistical mechanics.
Recently it was used for calculation of entanglement entropy in exactly solvable models including  spin chains. 
We use  Fisher-Hartwig formula  to calculate entanglement entropy [as well as R\'enyi entropy]
of large block of spins in the ground state of $XY$ spin chain. In the end of the paper we announce our recent results [with F. Franchini and L. A. Takhtajan] on  spectrum of  density matrix of the block of spins.
\end{abstract}

\section{Introduction}

We  study von Neumann entropy and  R\'enyi entropy  of spin chains  by means of the  { Fisher-Hartwig} formula.
The concept of  entanglement was introduced Schr\"odinger in 1935  in the course of developing the famous `cat paradox' , see \cite{S}.
Recently it became important  as a resource for quantum control, which is   central for quantum device building, including quantum computers (it is a primary resource for information processing).    Entropy of a subsystem as a measure of entanglement was introduced in  \cite{BBPS}.
We  study spin chains with unique ground state. Von Neumann entropy  (and R\'enyi entropy) of the whole ground state is zero, but it is positive for 
a subsystem [block of spins]. In order to define entanglement entropy one has to introduce reduced density matrix.
The reduced density matrix was first introduced by P. A. M. Dirac in 1930, see \cite{Dirac}. 

 We  calculate the entropy of a block of $L$ continuous spins in the ground state of a Hamiltonian. We can think that the ground state is a bipartite system $|GS\rangle=|A\& B\rangle$, where we call the block  by subsystem $A$ and the rest of the ground state by subsystem $B$. The density matrix of the ground state is $\rho_{AB}=|GS\rangle\langle GS|$, and the density matrix of the block of $L$ neighboring spins [subsystem $A$] is $\rho_{A}=\mathrm{Tr}_{B}\left(\rho_{AB}\right)$, where we trace out all degrees of freedom outside the block. The von Neumann entropy of the block is
\begin{eqnarray}
	S(\rho_{A})=-\mathrm{Tr}_{A}\left(\rho_{A}\ln \rho_{A}\right), \label{vnedef}
\end{eqnarray}
which measures how much the block is entangled with the rest of  the ground state. 
On the other hand, the { R\'enyi} entropy $S(\rho_{A}, \alpha)$ is defined as
\begin{eqnarray}
	S(\rho_{A}, \alpha)=\frac{1}{1-\alpha}\ln \mathrm{Tr}_{A} \left(\rho^{\alpha}_{A}\right), \qquad  \quad \mbox{and} \quad \alpha>0, \label{redef}
\end{eqnarray}
here $\alpha$ is a parameter .  R\'enyi entropy  \cite{renyi} is important in information theory.
 The R\'enyi entropy  turns into  von Neumann entropy at $\alpha\rightarrow 1$. Knowledge of the R\'enyi entropy at arbitrary $\alpha $ permits evaluation of  spectrum of the density matrix .
Our main example is $XY$ spin chain.

Toeplitz matrix \mbox{$T_{\mathrm{L}}[\Phi]$} is said to be
expressed in terms of  generating function \mbox{$\Phi(\theta)$} (which is
called symbol  in  mathematical literature):
\begin{equation}
T_{\mathrm{L}}[\Phi]= (\Phi_{i-j}),~~~i,j=1,\cdots,\mathrm{L}-1
\end{equation}
where
\begin{equation}\label{four_def}
\Phi_k=\frac{1}{2\pi} \int_0^{2\pi} \Phi (\theta) e^{-\mathrm{i} k \theta} \mathrm{d} \theta
\end{equation}
is the $k$-th Fourier coefficient of generating function
\mbox{$\Phi(\theta)$}. The generating function \mbox{$\Phi(\theta)$}
can be type of $N\times N$ matrix and \mbox{$T_{\mathrm{L}}[\Phi]$}
is a $N\, L\times N\, L$ matrix for such case. One of the central objects 
 in the study of Toeplitz matrix
\mbox{$T_{\mathrm{L}}[\Phi]$} is its determinant, which we will denote as
$D_{\mathrm{L}}[\Phi]$,
\begin{equation}\label{detdef}
D_{\mathrm{L}}[\Phi]:= \det T_{\mathrm{L}}[\Phi] .
\end{equation}

Starting with Onsager's celebrated solution of the two-dimensional Ising model in the 1940's, 
Toeplitz determinants play an increasingly central role in modern
mathematical physics. We refer the reader to the book  \cite{macwu}, 
and to  survey \cite{mccoy1} as for comprehensive sources of the  classical
results and the history concerning the use of Toeplitz determinants in statistical
mechanics. 

Another important areas of applications of Toeplitz determinants are random matrices 
and combinatorics. We refer the readers to the works \cite{tw,bdj,gessel}
for the basic results and for the historic reviews.

Given a generating function $\Phi(\theta)$, a principal question is the evaluation of  the large $L$ behavior 
of the Toeplitz determinant  $D_{\mathrm{L}}[\Phi]$. The pioneering 
works on the asymptotic analysis of Toelpitz determinants were done
by  Szeg\"o (regular symbol)   and by  Fisher and Hartwig (singular symbol).  These results have been used
in the study of spin correlation in two-dimensional Ising model in the classical
works of Wu and McCoy, see for example \cite{macwu} and since then  by many other
researchers and  for a  various generating functions. 

The main focus of the majority of  works in the area has been, so far,  the study of spin correlations.
The key objects of the analysis have been  the relevant correlation functions of the local
operators. In this paper, we discuss yet another,  more recent 
application of the asymptotic analysis of Toeplitz determinants in the theory
of quantum spin models. Instead of the local operators,
these applications are concerned with the important nonlocal objects 
appearing in spin chains in connection to their suggested use in quantum
informatics \cite{LRV}. Indeed, we shall survey  some of the recent results 
concerning the {\it quantum entanglement}. We will consider the two
applications - the entanglement in the XX  model and in the XY model. 
The first one is related to a singular scalar generating function, while the second
one deals with  a regular but  ( $2\times 2$) matrix generating function.

We begin with the brief review of the history and some of the most
recent results concerning the asymptotic analysis of Toeplitz determinants.

{  The {\bf plan} of the paper} is:

In the second section we discuss the  asymptotical expression of the determinant of a large Toeplitz matrix.
The section is divided into subsections. Subsection 2.2 is  devoted to block Toeplitz determinants.

Third section is devoted to $XY$ spin chain. In subsection 3.1 we remind derivation of determinant representation of entropy of 
a block of spins in the ground state. Isotropic case, i.e. the $XX$ model, is considered in 3.2. 
For anisotropic case we have to use block Toeplitz matrices.

In section 4 we derive asymptotic expression of entropy of large block of spins in isotropic case: the leading logarithmic term and sub-leading corrections. 

In section 5 we derive asymptotic expression of von Neuman entropy of large block of spins in anisotropic case.
In the case of $XY$ spin chain the entropy has a limit. We calculate the limit. 

In section 6 we calculate limiting expression for  Renyi entropy of large block of spins in $XY$ spin chain. 

In section 7 we  derive the spectrum of the limiting density matrix from Renyi entropy. We prove that the spectrum is exact
geometric sequence, see Eq. \ref{spectrumfinal}  and   Eq. \ref{spectrumfinal2}. We also calculate the degeneracy of individual eigenvalues, see f
Eq. \ref{asymp12}. 

The content of sections 4 - 7 is based on the works \cite{jin,ijk1,ijk2,fik,fikt}.

In section 8 we formulate open problems.

\section{Szeg\"o and Fisher-Hartwig Asymptotics}

Throughout the paper we will follow the usual, in the theory
and applications of Toeplitz determinants, convention to
denote the argument of the  functions on the unite circle either as 
$\theta$ or as $z$, $ z=e^{i\theta}$, i.e. we will always assume the
notational identity,
$$
f(z) \equiv f(\theta), \quad z = e^{\mathrm{i}\theta}, \quad \theta \in [0, 2\pi ).
$$
We first consider the case of scalar generating function, i.e. $N =1$. We shall also use 
for this case the low case symbol $\phi$ instead of $\Phi$. 

\subsection{Szeg\"o and Fisher-Hartwig asymptotics in the case of scalar symbols}

In this subsection we review the basic mathematical facts concerning  the asymptotics
of Toeplitz
determinants $D_{\mathrm{L}}[\phi]$ with scalar generation functions $\phi(z)$.

The large $L$  asymptotic behavior of  $D_{\mathrm{L}}[\phi]$ depends significantly 
on the analytical properties of the generating function $\phi(\theta)$. In the case of the smooth
enough functions  $\phi(\theta)$, the behavior is exponential and its leading 
and the pre-exponential terms are given by the following classical result of Szeg\"o,
known as the {\it  strong Szeg\"o limit theorem}.

\noindent
{\bf Theorem 1.}{\it Suppose that the generation function $\phi(\theta)$ satisfies the
conditions,
\begin{enumerate}
\item $\phi(\theta) \neq 0$ , for all $\theta \in [0, 2\pi )$.
\item $ \mathrm{index}\,\, \phi(\theta) \equiv \arg \phi(2\pi) - \arg\phi(0) = 0$
\item $\sum_{k=-\infty}^{\infty}|V_k| + \sum_{k=-\infty}^{\infty}|k||V_k|^2 < \infty$,
where $V_{k}$ are the Fourier coefficients  of the function,
\begin{equation}\label{szego3}
V(\theta) := \ln \phi(\theta),
\end{equation}
that is,
\begin{equation}\label{szego4}
\quad V(z)=\sum_{k=-\infty}^\infty V_k z^k,\qquad 
V_k={1\over 2\pi}\int_0^{2\pi}V(\th)e^{-ki\th}d\th.
\end{equation}
\end{enumerate}
Then, 
\begin{equation}\label{szego1}
D_{\mathrm{L}}[\phi]\sim  E_{\mathrm{Sz}}[\phi] \exp\Bigl( LV_{0}\Bigr),\quad L \to \infty,
\end{equation}
where the pre-exponential factor, $E_{\mathrm{Sz}}[\phi]$, is given by the equation{\footnote{
It is this equation which is responsible for the term ``strong Szeg\"o theorem''. Szeg\"o's first
result, i.e. {\it Szeg\"o limit theorem} produced the asymptotics of the determinant $D_{\mathrm{L}}[\phi]$
up to an undetermined multiplicative constant.}},
\begin{equation}\label{szego2}
E_{\mathrm{Sz}}[\phi] = \exp\Bigl(\sum_{k=1}^\infty k V_k V_{-k}\Bigr).
\end{equation}}

Conditions (1) and  (2) on the symbol $\phi(\theta)$ ensure that the function $V(z)$ is a well
defined function on  the unit circle. Condition (3) is a smoothness condition{\footnote{In \cite{Szego},
Szeg\"o proved this theorem under a somewhat stronger smoothness assumption on the symbol;
namely, he assumed that the symbol is positive, and that the symbol and its derivative are Lipshitz functions.
It took a substantial period of time and the efforts of several very skillful analysts to
reduce the smoothness conditions to the conditions  (1) - (2) above.  It also worth noticing that these conditions are
already precise, i.e.,  if they do not satisfy, the  asymptotics (\ref{szego1}) might not take place.}}.
It is certainly  satisfied by the differentiable functions and is not satisfied by the functions
having root and jump singularities. 
In the context of Toeplitz matrices, this
type of singularities is usually called the {\it Fisher-Hartwig singularities}. The
general form of the  symbol  
$\phi(z)$ which  has  $m$, $ m = 0, 1, 2,\dots$ fixed 
Fisher-Hartwig singularities  is given by the equation
{\footnote{In writing the Fisher-Hartwig symbol in form (\ref{fFH}) we follow the
recent paper \cite{dik}. Equation (\ref{fFH}) is slightly different from
the one accepted  in  most of the literature devoted to the
Fisher-Hartwig generating functions. The ``translation'' back
to the standard form is easy. The main deviation from the
standard form is that in (\ref{fFH}) the product $z^{\sum_{j=0}^m \bt_j}$
is  factored out which allow to better appreciate the non-trivialty of the
shifting some of the parameters $\beta_{j}$ by integers.}},
\be\la{fFH}
\phi(z)=e^{V(z)} z^{\sum_{j=0}^m \bt_j} 
\prod_{j=0}^m  |z-z_j|^{2\al_j}g_{z_j,\bt_j}(z)z_j^{-\bt_j},\qquad z=e^{i\th},\qquad
\theta\in[0,2\pi),
\ee
where
\begin{eqnarray}
&z_j=e^{i\th_j},\quad j=0,\dots,m,\qquad
0=\th_0<\th_1<\cdots<\th_m<2\pi;&\la{z}\\
&g_{z_j,\bt_j}(z)\equiv g_{\bt_j}(z)=
\left\{\begin{array}{rl}
e^{i\pi\bt_j}& 0\le\arg z<\th_j,\\
e^{-i\pi\bt_j}& \th_j\le\arg z<2\pi
\end{array}\right.,&\la{g}\\
&\Re\al_j>-1/2,\quad \bt_j\in\bbc,\quad j=0,\dots,m,&
\end{eqnarray}
and $V(z)$ is a sufficiently smooth function on the unit circle so that
the first factor of the right hand side of equation (\ref{fFH}) represents
the `` Szeg\"o part'' of the symbol.
The condition on $\al_j$ insures integrability.
As it has already been mentioned before,  a single Fisher-Hartwig singularity at $z_j$ consists of 
a root-type singularity
\be\la{za}
|z-z_j|^{2\al_j}=\left|2\sin\frac{\th-\th_j}{2}\right|^{2\al_j}
\ee
and a jump $g_{\bt_j}(z)$. 
A point $z_j$, $j=1,\dots,m$ is included in (\ref{z})
if and only if either $\al_j\neq 0$ or $\bt_j\neq 0$ (or both);
in contrast, we always fix $z_0=1$ even if $\al_0=\bt_0=0$ (note that 
$g_{\bt_0}(z)=e^{-i\pi\bt_0})$.
Observe that for each $j=1,\dots,m$, $z^{\beta_j} g_{\beta_j}(z)$ 
is continuous at $z=1$, and so for each $j$ 
each ``beta'' singularity produces a jump only at the point $z_j$.

In 1968,  M. Fisher and R. Hartwig \cite{fisher} suggested a formula for the
leading term of the asymptotic behavior for the Toeplitz determinant 
generated by the symbol (\ref{fFH}){\footnote{Some important partial results concerning the
asymptotics of the Toeplitz determinants with singular symbols were also 
obtained by A. Lenard \cite{lenard} and used by Fisher and Hartwig as 
a strong evidence in favor of their formula.}}. The principal insight of Fisher and Hartwig
was the observation that the  singularities of the symbol yield the appearance 
of the power-like factors in the asymptotics.  Indeed, in the case of all
$\beta_j = 0$,  the Fisher-Hartig formula reads
as follows.
\begin{equation}\label{fh1}
D_{\mathrm{L}}[\phi]\sim  E^{0}_{\mathrm{FH}}[\phi]L^{\sum_{j=0}^m\al_j^2} \exp\Bigl( LV_{0}\Bigr),\quad L \to \infty.
\end{equation}
The pre-exponential constant factor, $E^{0}_{\mathrm{FH}}[\phi]$, is more elaborated than 
its Szeg\"o counterpart $E_{\mathrm{Sz}}[\phi]$ from the Szeg\"o equation
(\ref{szego1}). The description of $E^{0}_{\mathrm{FH}}[\phi]$ involves 
a rather ``exotic'' special function - the Barnes' $G$ - function $G(x)$
which is defined by the equations (see e.g. \cite{ww}),
\begin{equation}\label{Barne}
G(1+x) = (2\pi)^{x/2} e^{-(x+1)x/2-\gamma_E x^2/2}
\prod_{n=1}^{\infty} \{ (1+x/n)^n e^{-x+x^2/(2n)}\},
\end{equation}
where $\gamma_E$ is Euler constant and its numerical value is
\mbox{$0.5772156649\cdots$}. 
The $G$ - function can be thought of as a ``discrete antiderivative" of the 
$\Gamma$ - function. The exact expression for 
$E^{0}_{\mathrm{FH}}[\phi]$ is given by the equation (cf.  Eq. \ref{szego2}),
$$
E^{0}_{\mathrm{FH}}[\phi]=\exp\left(\sum_{k=1}^\infty k V_k V_{-k}\right)
\prod_{j=0}^me^{\alpha_j\Bigl(V_0-V(z_j)\Bigr)}
$$
\begin{equation}\label{fh2}
\times
\prod_{0\le j<k\le m}
|z_j-z_k|^{-2\al_j\al_k}\prod_{j=0}^m\frac{G^2(1+\al_j)}{G(1+2\al_j)}.
\end{equation}
The double product over $j<k$ is set to $1$
if $m=0$, so that in the absence of singularities, 
we are back to the strong Szeg\"o limit  theorem.

Fisher-Hartwig formula (\ref{fh1}) was proven in 1973 by H. Widom \cite{W}.

The presence of  jumps,
under the assumption $|\Re \beta_j - \Re \beta_k | < 1$, does not change much the 
structure of the large $L$ behavior of the Toeplitz  determinant $D_L[\phi]$. Indeed,
it is still  the combination of the exponential and the power terms with the exponential
term being determined, as before, by only the Szeg\"o part of the symbol while the
power factor is determined by both the $\alpha$ and  the $\beta$ parameters 
of the Fisher-Hartwig part of the symbol. The Fisher-Hartwig  formula for the
general case of symbol (\ref{fFH}) reads (cf. Eq. \ref{fh1}),
\begin{equation}\label{fh3}
D_{\mathrm{L}}[\phi]\sim  E_{\mathrm{FH}}[\phi]L^{\sum_{j=0}^m(\al_j^2-\bt_j^2)} \exp\Bigl( LV_{0}\Bigr),\quad L \to \infty.
\end{equation}
The pre-exponential constant factor, $E_{\mathrm{FH}}[\phi]$, is now even  more complex than 
in the case of  all $\beta_j = 0$. In addition to  the Barnes' $G$ - function, it now involves 
the canonical Wiener-Hopf factorization of the Szeg\"o part,  $e^{V(z)}$,
of the symbol $\phi(z)$, 
\be\la{WienH}
e^{V(z)}=b_+(z) e^{V_0} b_-(z),\qquad b_+(z)=e^{\sum_{k=1}^\infty V_k z^k},
\qquad b_-(z)=e^{\sum_{k=-\infty}^{-1} V_k z^k}.
\ee
Note that $b_+(z)$ and  $b_-(z)$ are analytic inside and outside of the unit circle $|z| =1$,
respectively, and they
satisfy the normalization conditions $b_+(0) = b_-(\infty) = 1$. The exact expression for 
$E_{\mathrm{FH}}[\phi]$ is given by the equation (cf. Eq. \ref{szego2} and Eq. \ref{fh2}),
$$
E_{\mathrm{FH}}[\phi]=\exp\left(\sum_{k=1}^\infty k V_k V_{-k}\right)
\prod_{j=0}^m b_+(z_j)^{-\al_j+\bt_j}b_-(z_j)^{-\al_j-\bt_j}
$$
$$
\times
\prod_{0\le j<k\le m}
|z_j-z_k|^{2(\bt_j\bt_k-\al_j\al_k)}\left({z_k\over z_j e^{i\pi}}
\right)^{\al_j\bt_k-\al_k\bt_j}
$$
\begin{equation}\label{fh4}
\times
\prod_{j=0}^m\frac{G(1+\al_j+\bt_j) G(1+\al_j-\bt_j)}{G(1+2\al_j)}
\left(1+o(1)\right).
\end{equation}

The proof of the general Fisher-Hartwig formula (\ref{fh3}) 
 is due to E. Basor \cite{B}  for $\Re \beta_j =0$, E. Basor \cite{B2}
for $\alpha_j =0$, $|\Re\beta_j| < 1/2$,
A. B\"ottcher and B. Silbermann \cite{BS} for $|\Re\alpha_j| < 1/2$,
$|\Re\beta_j| < 1/2$,  T. Ehrhardt \cite{Ehr} for $|\Re \beta_j - \Re \beta_k | < 1$.
The precise statement concerning the large $L$ behavior of the Toeplitz determinant 
$D_{L}[\phi]$ with the Fisher-Hartwig generating function (\ref{fFH})  is given by the following theorem.

\noindent
{\bf Theorem 2.} (T. Ehrhardt  \ci{Ehr}) {\it 
Let $\phi(z)$ be defined in (\ref{fFH}), $V(z)$ be $C^\infty$ on the unit circle,
$\Re\al_j>-1/2$, $|\Re\bt_j-\Re\bt_k|<1$, and $\al_j\pm\bt_j\neq -1,-2,\dots$
for $j,k=0,1,\dots,m$. Then, as $L\to\infty$, the asymptotic behavior of the
Toeplitz determinant $D_{L}[\phi]$ is given by the formulae (\ref{fh3}) -  (\ref{fh4}).}

A. B\"ottcher and B. Silbermann \cite{BS} in 1985 and E. Basor and C. Tracy  \cite{BT}
in 1991 constructed examples with $\Re \beta_{j}$ not lying in a single interval of length 
less than $1$  and such that the large $L$ asymptotics is very different
from the one given by (\ref{fh3}). These examples have  showed that for the asymptotics (\ref{fh3})
to take place,  the condition
\begin{equation}\label{betacond}
|\Re \beta_j - \Re \beta_k | < 1, \quad \forall j, k = 0, 1, ..., m,
\end{equation}
is precise. In the case of arbitrary complex $\beta_j$, E. Basor and C. Tracy 
conjectured in \cite{BT} a very elegant structure of the large $L$ asymptotics 
of the determinant $D_{L}[\phi]$. They based their arguments
on the formal analysis of the behavior of the both sides of estimate  (\ref{fh3})
with respect to the shifts of the $\beta$ - parameters by integers. A detail description
of the Basor-Tracy conjecture can be found in the original paper \cite{BT} as well
as in the recent work \cite{dik} were this conjecture was actually proven with the help
of the new technique - the {\it Riemann-Hilbert method}.

We refer the reader to  monograph \cite{abbs} and  survey \cite{Ehr} for
more on the mathematics of Toeplitz determinants with the Fisher-Hartwig
symbols.

For the Riemann-Hilbert approach in the theory of Toepitz determinants, we refer
the reader to the papers \cite{dik} and \cite{D} where the method was introduced 
(following the similar approach for the Hankel determinants \cite{fokik} 
and the theory of integrable Fredholm determinants \cite{iiks})
and to the  works   \cite{krasovsky1,krasovsky2,MMS1,MMS2}, 
where the method was further developed. The crucial role in the implementation
of the Riemann-Hilbert approach to the Toeplitz determinants  is played by the Deift-Zhou nonlinear steepest descent method
of the asymptotic analysis of the oscillatory matrix Riemann-Hilbert problems \cite{DZ}
and by its orthogonal polynomial version \cite{Dstrong}.

\subsection{Block Toeplitz determinants}

A general asymptotic representation of the determinant of a block Toeplitz matrix, which
generalizes the classical strong Szeg\"o theorem to the block matrix
case, was obtained by Widom in \cite{widom,widom2} (see also more recent
work \cite{bottcher} and references therein). 

\noindent
{\bf Theorem 4.} (H. Widom \cite{widom2}) {\it Let $\Phi(z)$ be a $N\times N$ matrix function
defined on the unit circle and satisfying the conditions, 
\begin{enumerate}
\item $\det\Phi(\theta) \neq 0$ , for all $\theta \in [0, 2\pi )$.
\item $ \mathrm{index}\,\, \det \Phi(\theta) \equiv \arg \det \Phi(2\pi) - \arg\det\Phi(0) = 0$
\item $\sum_{k=-\infty}^{\infty}|\Phi_k| + \sum_{k=-\infty}^{\infty}|k||\Phi_k|^2 < \infty$,
\end{enumerate}
where $\Phi_{k}$ are the Fourier coefficients of $\Phi(\theta)$, and $|F|$ denote a matrix
norm of the matrix F. Then, the asymptotic behavior of the block Topelitz determinant generated by
the symbol $\Phi(z)$ is given by the formulae,
\begin{equation}\label{widom1}
D_{\mathrm{L}}[\Phi]\sim  E_{\mathrm{W}}[\Phi] \exp\left( 
\frac{L}{2\pi} \int_0^{2\pi} \ln\det\Phi (\theta)\mathrm{d} \theta\right),\quad L \to \infty,
\end{equation}
\begin{equation}\label{widom2}
E_{\mathrm{W}}[\Phi] = \det\Bigl(T_{\infty}[\Phi]T_{\infty}[\Phi^{-1}]\Bigr).
\end{equation}
where $T_{\infty}[\Phi]$ is a semi-infinite Toeplitz matrix,
\begin{equation}\label{widom3}
T_{\infty}[\Phi]= (\Phi_{i-j}),~~~i,j=1, 2, \cdots.
\end{equation}}
From the application point of view, there is an important difference between 
this result and  the Szeg\"o formula (\ref{szego1}) for the case  of scalar symbols. 
Indeed, the determinant in the right hand side of Eq. \ref{widom2} is the Fredholm determinant
of an infinite matrix, and already for  $2 \times 2$ matrix symbols  the question 
of  effective evaluation  of Widom's pre-factor $E_{\mathrm{W}}[\Phi]$  is a highly 
nontrivial one,  even for a relatively simple matrix functions $\Phi$. Indeed, up until very recently
the only general class of matrix  functions $\Phi$ for which  such effective evaluation is possible
has been the class of functions with at least one-side truncated Fourier series.
This class was singled out by Widom himself in \cite{widom}, and this Widom's
result  has been used in the recent paper \cite{BE5} of E. Basor and T. Ehrhardt devoted 
to the dimer model.

Another class of matrix generating functions which admits an explicit evaluation of 
Widom's constant are the algebraic symbols. This fact was demonstrated in the
works \cite{ijk1,ijk2,imm} for important cases of the block Toeplitz determinants
appearing in the analysis of the entanglement  entropy in quantum  spin chains. For this
class of symbols, Widom's pre-factor admits an explicit evaluation in terms of
Jacobi and  Riemann theta functions. To give a flavor of these results, we 
will now present a detail description of the asymptotics of the block Toeplitz determinant 
related to the XY spin model obtained in \cite{ijk1,ijk2}. We shall also use these formulae 
later in Section 4.

The Toeplitz determinant in question is generated by the $2\times 2$ matrix symbol,
\begin{equation}\label{block1}
\Phi(z) = 
\left(\begin{array}{cc}
               \mathrm{i}\lambda & \phi(z)\\
               -\phi^{-1}(z)&\mathrm{i}\lambda
               \end{array} \right) \label{defphi0}
\end{equation}
\begin{equation}
\textrm{and}\quad \phi(z)=\sqrt{\frac{(z-z_1)(z-z_2)}{(1-z_{1}z)(1-z_{2}z)}},
\label{defph0}
\end{equation}
where $z_1 \neq z_2$ are complex nonzero numbers not lying on the unit circle. Following
the needs of the XY model, we shall assume that the both points are from the right half plane
though the result we present below can be easily generalized to the arbitrary position
of the points $z_1$ and $z_2$ outside of the unit  circle. We will also
distinguish three possible  locations of the points $z_1$ and $z_2$ on complex
plane.

\noindent
{\bf Case 1}a : Both $z_1$ and $z_2$ are real, they lie outside of the unit  circle, and we assume that $z_1 > z_2 >1$.

\noindent
{\bf Case 1}b : Both $z_1$ and $z_2$ are complex, $z_1 = z^{*}_2$, and we assume that
 $\Re z_1 >1$ and $\Im z_1 > 0$. 
 
 \noindent
{\bf Case 2} : Both $z_1$ and $z_2$ are real, they lie at the different sides of the unit circle, 
and we assume that $z_1 > z^{-1}_2 >1$.

\noindent
The reason why the Cases 1a and 1b are considered as sub-cases of a single case is
that in the both these cases all the root singularities of the function $\phi(z)$ defined
in (\ref{defph0}) are inside of
the unite circle while all its zeros are outside. In Case 2, the zeros and the singularities 
are evenly distributed between the inside and the outside of the unit circle. This difference
in the position of the roots and singularities of $\phi(z)$ has an impact to the derivations
of the asymptotics and, as we see below, is reflected in the form of the final answer.
We shall also see that in the context of the  XY model,   Case 1 and Case 2 correspond to the small ($h < 2$) and 
large ($h > 2$) magnetic field, respectively.

The complex  parameter $\lambda$ plays role of a spectral parameter for the 
Toeplitz matrix generated by the symbol,
\begin{equation}\label{block2}
\Phi_{0}(z) \equiv - \Phi(z)|_{\lambda = 0} =  
\left(\begin{array}{cc}
               0 & -\phi(z)\\
               \phi^{-1}(z)& 0
               \end{array} \right) \label{defphi01}.
\end{equation}
Hence the Toeplitz determinant $D_{L}[\Phi]$ we are dealing with is in fact 
a Toeplitz {\it characteristic} determinant,
\begin{equation}\label{chracter}
D_{L}[\Phi] \equiv D_{L}(\lambda) = \det \Bigl(i\lambda I_{2L} - T_{\L}[\Phi_0]\Bigr).
\end{equation}

Given the branch points $z_j$ of the symbol $\Phi(z)$, we  introduce now the elliptic curve,
\begin{equation}\label{curve}
w^{2}(z) = (z-z_1)
(z-z_2)(z-z_2^{-1})(z-z_1^{-1}).
\end{equation}
Let us also re-label the branch points of this curve by the letters $\lambda_A$,
$\lambda_B$, $\lambda_C$, and $\lambda_D$, according to the following rule.
{\bf Case 1}a : $\lambda_A = z^{-1}_{1}$, $\lambda_B = z^{-1}_{2}$,
$\lambda_C = z_{2}$, $\lambda_D= z_1$;
{\bf Case 1}b : $\lambda_A = z^{-1}_{1}$, $\lambda_B = z^{-1}_{2}$,
$\lambda_C = z_{1}$, $\lambda_D= z_2$,
{\bf Case 2} : $\lambda_A = z^{-1}_{1}$, $\lambda_B = z_{2}$,
$\lambda_C = z^{-1}_{2}$, $\lambda_D= z_1$.
Observe that $\lambda_A$ and $\lambda_B$ are always inside the
unite circle while $\lambda_C$ and $\lambda_D$ are always outside.
This new relabeling of the branch points allows to introduce the
module parameter of elliptic curve (\ref{curve}) in the universal way,
\begin{equation}
\tau=\frac{2}{c} \int_{\lambda_B}^{\lambda_C}\frac{\mathrm{d}
z}{w(z)}, \quad
c=2\int_{\lambda_A}^{\lambda_B}\frac{\mathrm{d}z}{w(z)}.\label{module}
\end{equation}

\noindent
{\bf Theorem 5.} (\cite{ijk1,ijk2}) 
{\it Let 
\begin{equation}
\theta_3(s)=\sum_{n=-\infty}^{\infty} e^{\pi i \tau n^2+2\pi i s
n},\label{jac}
\end{equation}
where $\tau$ is taken from (\ref{module}), be the third Jacobi theta-function
associated with the curve  (\ref{curve}). 
Then,  the large $L$ asymptotic behavior of the determinant
$D_{L}(\lambda)$ is given by the equations,
\begin{equation}\label{block3}
D_L(\lambda) \sim \frac{ \theta_{3}\left( \beta(\lambda)+
\frac{\sigma\tau}{2}\right) \theta_{3}\left(\beta(\lambda) -
\frac{\sigma\tau}{2}\right)}{\theta^{2}_{3}\left(\frac{\sigma\tau}{2}\right)}
(1 - \lambda^2)^{L}, \quad L \to \infty
\end{equation}
where 
\begin{equation}\label{betablock}
\beta(\lambda) = \frac{1}{2\pi i}\ln\frac{\lambda+1}{\lambda-1},
\end{equation}
and $\sigma = 1$ in Case 1 and $\sigma = 0$ in Case 2.}

\noindent
{\bf Remark.} 
The theta-functions involved in the asymptotic formula Eq. (\ref{block3})
has zeros at the points 
\begin{equation}\label{zeros}
 \pm \lambda_{m}, \quad \lambda_{m} = \tanh \left(m +
\frac{1-\sigma}{2}\right)\pi \tau_{0}, \quad m \geq 0,
\end{equation}
where,
$$
\tau_{0} = -i\tau = -i
\frac{\int_{\lambda_{B}}^{\lambda_{C}}\frac{dz}{w(z)}}
{\int_{\lambda_{A}}^{\lambda_{B}}\frac{dz}{w(z)}}>0 .
$$
The asymptotics  (\ref{block3}) is uniform  outside of the arbitrary fixed
neighborhoods of the points $\lambda = \pm1$ and $\lambda = \pm \lambda_m$.

Observe that in the case under consideration, $\det \Phi(z) \equiv 1 -\lambda^2$. Therefore, the last factor in (\ref{block3})
is exactly the exponential term of the general Widom-Szeg\"o formula (\ref{widom1})
written for  symbol (\ref{defphi0}). The rest of  (\ref{block3}) gives then the corresponding
Widom's constant, i.e.
\begin{equation}\label{widomXY}
E_{W}[\Phi] = \frac{ \theta_{3}\left( \beta(\lambda)+
\frac{\sigma\tau}{2}\right) \theta_{3}\left(\beta(\lambda) -
\frac{\sigma\tau}{2}\right)}{\theta^{2}_{3}\left(\frac{\sigma\tau}{2}\right)}.
\end{equation}

Similar formulae for the case of the more general quantum spin chains were obtained in \cite{imm}.
The relevant generating function has the same matrix structure (\ref{defphi0}) with the scalar
function $\phi(z)$ defined by the equation,
\begin{equation}
  \label{eq:g_def}
   \phi(z) := \sqrt{\frac{p(z)}{z^{2n}p(1/z)}}
\end{equation}
and  $p(z)$ is a polynomial of degree $2n$.  The analog of the formulae (\ref{block3}) - (\ref{widomXY})
in the case $n >1$ involves, instead of elliptic, the hyperelliptic integrals and, instead of the Jocobi
theta-function, the $2n -1$ dimensional Riemann theta-function.

The methods that lead to these results,
involves the theory of integrable Fredholm operators \cite{iiks,hi,D}
and the use of the algebrageometric techniques of the soliton theory (see e.g. \cite{bbeim}).

\section{$XY$ Model and Block Entropy}

The Hamiltonian of $XY$  model can be written as
\begin{equation}
H=-\sum_{n=-\infty}^{\infty}
(1+\gamma)\sigma^x_{n}\sigma^x_{n+1}+(1-\gamma)\sigma^y_{n}\sigma^y_{n+1}
+ h\sigma^z_{n} \label{xxh}
\end{equation}
Here $\sigma^x_n$, $\sigma^y_n$ $\sigma^z_n$ are Pauli matrices  and
$h$ is a magnetic field; Without loss generality, the anisotropy
parameter $\gamma$ can be taken as $0\le\gamma$; Case with
$\gamma=0$ is usually called $XX$ model. The model was solved in
\cite{Lieb,gallavotti,mccoy,mccoy2} and it owns a unique ground
state $|GS\rangle$. Toeplitz determinants were used for evaluation
of some correlation functions\cite{mccoy2,aban}; Integrable Fredholm
operators were used for calculation of other correlations\cite{sla,dz,pron}.
 When the system is in the ground state, the entropy for this whole system is zero but
the entropy of a sub-system can be positive. We calculate the
entropy of a sub-system (a block of $\mathrm{L}$ neighboring spins)
which can measure the entanglement between this sub-system and the
rest part\cite{jin}. We treat the whole chain as a binary system
$|GS\rangle = |A \& B\rangle $, where we denote the block of $L$
neighboring spins by sub-system A and the rest part by sub-system B.
The density matrix of the ground state can be denoted by
\mbox{$\rho_{AB}=|GS\rangle \langle GS|$}. The density matrix of
sub-system A is \mbox{$\rho_A= Tr_B(\rho_{AB})$}. Von Neumann
entropy $S(\rho_A)$  of the sub-system A can be represented as
following:
\begin{equation}
S(\rho_A)=-Tr_A(\rho_A \ln \rho_A). \label{edif} \label{olds}
\end{equation}
This entropy also defines the dimension of the Hilbert space of
states of the block of $L$ spins.

\subsection{Derivation}
 Following Ref.~\cite{Lieb}, \cite{kauf}, we  introduce two Majorana
operators
\begin{equation}
c_{2l-1}= (\prod_{n=1}^{l-1} \sigma^z_{n}) \sigma^x_l~~~\textrm{and}
~~~c_{2l}= (\prod_{n=1}^{l-1} \sigma^z_{n}) \sigma^y_l,
\end{equation}
on each site of the spin chain. Operators $c_n$ are hermitian and
obey the anti-commutation relations \mbox{$\{ c_m, c_n\} =2
\delta_{mn}$}. In terms of operators $c_n$, Hamiltonian
$H_{\mathrm{XX}}$ can be rewritten as
\begin{equation}
H_{\mathrm{XX}}(h)= i \sum_{n=1}^N (c_{2n}c_{2n+1}-
c_{2n-1}c_{2n+2}+ h c_{2n-1}c_{2n}).
\end{equation}
Here different boundary effects can be ignored because we are only
interested in cases with \mbox{$N\to \infty$}. This Hamiltonian can
be subsequently diagonalized by linearly transforming the operators
$c_n$. It has been obtained \cite{Lieb,mccoy} (also see
\cite{vidal,LRV}) that
\begin{equation}
\langle GS| c_m|GS\rangle=0,~~\langle GS|c_m c_n|GS\rangle
=\delta_{mn}+i (\mathbf{B}_N)_{mn}.
\end{equation}
Here matrix $\mathbf{B}_N$ can be written in a block form as
\begin{equation}
\mathbf{B}_N=\left( \begin{array}{cccc}
\Pi_0 &\Pi_{-1}& \ldots &\Pi_{1-N}\\
\Pi_{1}& \Pi_0&   &   \vdots\\
\vdots &      & \ddots&\vdots\\
\Pi_{N-1}& \ldots& \ldots& \Pi_0
\end{array}     \right) \quad \textrm{and}\quad
\Pi_l=\frac{1}{2\pi} \int_{0}^{2\pi} \, \mathrm{d} \theta\,
e^{-\mathrm{i} l \theta} {\Phi}_{0}(\theta) \label{bn},
\end{equation}
where both $\Pi_l$ and $\Phi_0(\theta)$ (for \mbox{$N \to
\infty$}) are \mbox{$2\times 2$} matrix,
\begin{equation}
\Phi_0(\theta)=\left( \begin{array}{cc}
               0& \phi(\theta)\\
               -\phi^{-1}(\theta)&0
               \end{array} \right)\quad \textrm{and} \quad \phi(\theta)=\frac{\cos \theta -\mathrm{i}
\gamma\sin \theta -h/2}{|\cos \theta -\mathrm{i} \gamma\sin \theta
 -h/2|}.
\end{equation}
Other correlations such as \mbox{$\langle GS|c_m \cdots
c_n|GS\rangle$} are obtainable by Wick theorem. The Hilbert space of
sub-system A can be spanned by \mbox{$\prod_{i=1}^{\mathrm{L}}
\{\sigma^{-}_i\}^{p_i}|0\rangle_F$}, where $\sigma^{\pm}_i$ is Pauli
matrix, $p_i$ takes value $0$ or $1$, and vector
\mbox{$|0\rangle_F$} denotes the ferromagnetic state with all spins
up. It's possible to construct a set of fermionic operators $b_i$
and $b^{+}_i$ by defining
\begin{equation}
d_m= \sum_{n=1}^{2\mathrm{L}} v_{mn} c_n,~~m=1,\cdots,
2\mathrm{L};~~~ b_l= (d_{2l} +i d_{2l+1})/2, ~~l=1,\cdots,
\mathrm{L}\label{ctob}
\end{equation}
with \mbox{$v_{mn}\equiv (\mathbf{V})_{mn}$}. Here matrix
$\mathbf{V}$ is an orthogonal matrix. It's easy to verify that $d_m$
is hermitian operator and
\begin{equation}
b^+_l= (d_{2l} -i
d_{2l+1})/2,~~~\{b_i,b_j\}=0,~~~\{b^{+}_i,b^{+}_j\}=0,~~~\{b^{+}_i,b_j\}=\delta_{i,j}.
\end{equation}
In terms of fermionic operators $b_i$ and $b^{+}_i$, the Hilbert
space can also be spanned by \mbox{$\prod_{i=1}^{\mathrm{L}}
\{b^{+}_i\}^{p_i}|0 \rangle_{vac}$}. Here $p_i$ takes value $0$ or
$1$, $2\mathrm{L}$ fermionic operators $b_i$, $b^{+}_i$ and vacuum
state \mbox{$|0\rangle_{vac}$} can be constructed by requiring
\begin{equation}
b_l|0\rangle_{vac}=0,~~l=1,\cdots, \mathrm{L}.
\end{equation}
We shall choose a specific orthogonal matrix $\mathbf{V}$ later.

Let $\{\psi_I\}$ be a set of orthogonal basis for Hilbert space of
any physical system. Then the most general form for density matrix
of this physical system can be written as
\begin{equation}
\rho=\sum_{I,J} c(I,J) |\psi_I\rangle \langle\psi_J|.
\end{equation}
Here \mbox{$c(I,J)$} are complex coefficients. We can introduce a
set of operators \mbox{$P(I,J)$} by $ P(I,J) \propto |\psi_I\rangle
\langle\psi_J|$ and \mbox{$\widetilde{P}(I,J)$} satisfying
\begin{equation}
\widetilde{P}(I,J) P(J,K)=\delta_{I,K} |\psi_I\rangle
\langle\psi_I|,~~ P(I,J) \widetilde{P}(J,K)=\delta_{I,K}
|\psi_I\rangle \langle\psi_I|.
\end{equation}
There is no summation over repeated index in these formula. We shall
use an explicit summation symbol through the whole paper. Then we
can write the density matrix as
\begin{equation}
\rho=\sum_{I,J} \tilde{c}(I,J) P(I,J),~~ \tilde{c}(I,J)=Tr(\rho
\widetilde{P}(J,I)).
\end{equation}
Now let us consider quantum spin chain defined in Eq.~$\ref{xxh}$.
For the sub-system A, the complete set of operators \mbox{$P(I,J)$}
can be generated by \mbox{$\prod_{i=1}^{\mathrm{L}} O_i$}. Here we
take operator $O_i$ to be any one of the four operators $\{ b^{+}_i,
b_i, b^{+}_i b_i, b_i\, b^{+}_i\}$ (Remember that $b_i$ and $b_i^+$
are fermionic operators defined in Eq.~\ref{ctob}). It's easy to
find that \mbox{$\widetilde{P}(J,I)=(\prod_{i=1}^{\mathrm{L}}
O_i)^{\dagger}$}  if \mbox{$P(I,J)=\prod_{i=1}^{\mathrm{L}} O_i$}.
Here ${\dagger}$ means hermitian conjugation. Therefore, the reduced
density matrix for sub-system A can be represented as
\begin{equation}
\rho_A=\sum Tr_{AB} \left(\rho_{AB} (\prod_{i=1}^{\mathrm{L}}
O_i)^{\dagger}\right) \prod_{i=1}^{\mathrm{L}} O_i.
\end{equation}
Here the summation is over all possible different terms
\mbox{$\prod_{i=1}^{\mathrm{L}} O_i$}. For the whole system to be in
pure state \mbox{$|GS\rangle$}, the density matrix $\rho_{AB}$ is
represented by \mbox{$|GS\rangle \langle GS|$}. Then we have the
expression for $\rho_A$ as following
\begin{equation}
\rho_A=\sum \langle GS|(\prod_{i=1}^{\mathrm{L}} O_i)^{\dagger}|GS
\rangle \prod_{i=1}^{\mathrm{L}} O_i\;. \label{ll19}
\end{equation}
This is the expression of density matrix with the coefficients
related to multi-point correlation functions. These correlation
functions are well studied in the physics literature \cite{korepin}.
Now let us choose matrix $\mathbf{V}$ in Eq.~$\ref{ctob}$ so that
the set of fermionic basis $\{ b^+_i\}$ and $\{ b_i\}$ satisfy an
equation
\begin{equation}
\langle GS| b_i b_j|GS \rangle= 0,~~ \langle GS| b^+_i b_j|GS
\rangle= \delta_{i,j} \langle GS|b^+_i b_i|GS\rangle. \label{ddb}
\end{equation}
Then the reduced density matrix $\rho_A$ represented as sum of
products in Eq.~$\ref{ll19}$ can be represented as a product of sums
\begin{equation}
\rho_A= \prod_{i=1}^{\mathrm{L}} \Bigl( \langle GS|b^+_i b_i|GS
\rangle b^+_i b_i+\langle GS|b_i b^+_i|GS \rangle b_i b^+_i
\Bigr).\label{dmf}
\end{equation}
Here we used the equations \mbox{$\langle GS|
b_i|GS\rangle=0=\langle GS| b^+_i|GS\rangle$} and Wick theorem. This
fermionic basis was suggested  in Ref.~\cite{vidal,LRV}.

Now let us find a matrix $\mathbf{V}$ in Eq.~$\ref{ctob}$, which
will block-diagonalize correlation functions of Majorana operators
$c_n$. From Eqs.~$\ref{ctob}$ and $\ref{bn}$, we have the following
expression for correlation function of $d_n$ operators:
\begin{eqnarray}
\langle GS|d_m d_n|GS\rangle&=&\sum_{i=1}^{2\mathrm{L}} 
\sum_{j=1}^{2\mathrm{L}} v_{mi} \langle GS|c_i c_j|GS\rangle v_{jn} \;,\nonumber\\
\langle GS|c_m c_n|GS\rangle&=&\delta_{mn} +\mathrm{i} (\mathbf{B}_{\mathrm{L}})_{mn},\nonumber\\
\langle GS|d_md_n|GS\rangle&=&\delta_{mn}+ \mathrm{i}
(\widetilde{\mathbf{B}}_{\mathrm{L}})_{mn}.\label{diaB}
\end{eqnarray}
The last equation is the definition of a matrix
$\widetilde{\mathbf{B}}_{\mathrm{L}}$. Matrix
$\mathbf{B}_{\mathrm{L}}$ is the sub-matrix of
$\mathbf{B}_{\mathrm{N}}$ defined in Eq.~\ref{bn} with
$m,n=1,2,\dots,\mathrm{L}$. We also require
$\widetilde{\mathbf{B}}_{\mathrm{L}}$ to be the form
\cite{vidal,LRV}
\begin{equation}
\widetilde{\mathbf{B}}_{\mathrm{L}}=V \mathbf{B}_{\mathrm{L}} V^T=
\oplus_{m=1}^{\mathrm{L}} \nu_m \left( \begin{array}{cc}
               0& 1\\
               -1&0
               \end{array} \right)= \mathbf{\Omega}\otimes \left( \begin{array}{cc}
               0& 1\\
               -1&0
               \end{array} \right).\label{mmf}
\end{equation}
Here matrix $\mathbf{\Omega}$ is a diagonal matrix with elements
$\nu_m$ (all $\nu_m$ are real numbers). Therefore, choosing matrix
$\mathbf{V}$ satisfying Eq.~$\ref{mmf}$ in Eq.~$\ref{ctob}$, we
obtain $2\mathrm{L}$ operators $\{ b_l\}$ and $\{ b^+_l\}$ with
following expectation values
\begin{equation}
\langle GS|b_m|GS \rangle=0\;,\langle GS|b_m b_n|GS\rangle
=0\;,\langle GS|b^+_m b_n|GS \rangle =\delta_{mn}\frac{1+\nu_m}{2}.
\end{equation}
Using the simple expression for reduced density matrix $\rho_A$ in
Eq.~$\ref{dmf}$, we obtain
\begin{equation}
\rho_A= \prod_{i=1}^{\mathrm{L}} \left( \frac{1+\nu_i}{2} b^+_i b_i+
\frac{1-\nu_i}{2} b_i b^+_i \right).
\end{equation}
This form immediately gives us all the eigenvalues
$\lambda_{x_1x_2\cdots x_{\mathrm{L}}}$ of reduced density matrix
$\rho_A$,
\begin{equation}
\lambda_{x_1x_2\cdots x_{\mathrm{L}}}=\prod_{i=1}^{\mathrm{L}}
\frac{1+(-1)^{x_i} \nu_i}{2},~~~x_i=0,1~~\forall i. \label{eigf}
\end{equation}
Note that in total we have $2^{\mathrm{L}}$ eigenvalues. Hence, the
entropy of $\rho_A$ from Eq.~$\ref{edif}$ becomes
\begin{equation}
S(\rho_A)=\sum_{m=1}^{\mathrm{L}} e(1,\nu_m) \label{eaap}
\end{equation}
with
\begin{equation}
 e(x, \nu)= -\frac{x+\nu}{2} \ln (\frac{x+\nu}{2})-\frac{x-\nu}{2} \ln (\frac{x-\nu}{2}).\label{intee}
\end{equation}

\subsection{$XX$ model}

Notice further that for $XX$ model, i.e. $\gamma=0$ case, matrix
$\mathbf{B}_{\mathrm{L}}$ can have a direct product form
\begin{equation}
\mathbf{B}_{\mathrm{L}}= \mathbf{G}_{\mathrm{L}} \otimes \left(
\begin{array}{cc}
0& 1\\
-1&0
\end{array} \right)\quad \label{bgm}\textrm{with} \quad\mathbf{G}_{\mathrm{L}}=\left( \begin{array}{cccc}
\phi_0 &\phi_{-1}& \ldots &\phi_{1-L}\\
\phi_{1}& \phi_0&   &   \vdots\\
\vdots &      & \ddots&\vdots\\
\phi_{L-1}& \ldots& \ldots& \phi_0
\end{array}     \right)\;, \label{mg1}
\end{equation}
where $\phi_l$ is defined as
\begin{equation}
\phi_l=\frac{1}{2\pi} \int_0^{2\pi} \, \mathrm{d} \theta\,
e^{-\mathrm{i} l \theta} \phi(\theta), \quad \phi(\theta)=\left\{
\begin{array}{rl}
1, & -k_F < \theta <k_F,\\
-1, & k_F < \theta < (2\pi-k_F)
\end{array} \right.\label{mg2}
\end{equation}
and \mbox{$k_F=\arccos(|h|/2)$}. From Eqs.~$\ref{mmf}$ and
$\ref{bgm}$, we conclude that all $\nu_m$ are just the eigenvalues
of real symmetric matrix $\mathbf{G}_{\mathrm{L}}$.

\begin{figure}[ht]
\begin{center}
\includegraphics[width=3in,clip]{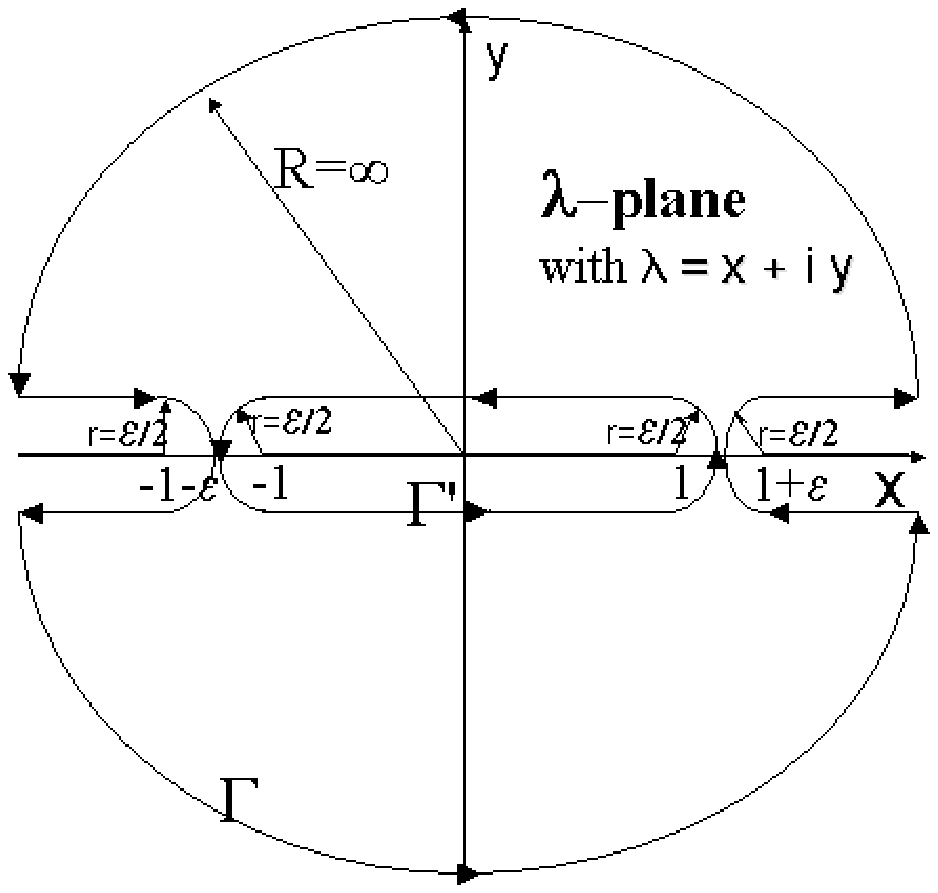}
\end{center}
\caption{\it Contours \mbox{$\Gamma'$} (smaller one) and
\mbox{$\Gamma $} (larger one). Bold lines $(-\infty, -1-\epsilon)$
and $(1+\epsilon,\infty)$ are the cuts of integrand
$e(1+\epsilon,\lambda)$. Zeros of $D_{L}(\lambda)$ (Eq.~$\ref{exd}$)
are located on bold line $(-1, 1)$. The arrow is the direction of
the route of integral we take and $\mathrm{r}$ and $\mathrm{R}$ are
the radius of circles. $\P $  } \label{fig1}
\end{figure}

However, to obtain all eigenvalues $\nu_m$ directly from matrix
$\mathbf{G}_{\mathrm{L}}$ is a non-trivial task. Let us introduce
\begin{equation}
D_{\mathrm{L}}(\lambda)= \det
(\widetilde{\mathbf{G}}_{\mathrm{L}}(\lambda) \equiv \lambda
I_{\mathrm{L}}- \mathbf{G}_{\mathrm{L}})\;.
\end{equation}
Here $\widetilde{\mathbf{G}}_{\mathrm{L}}$ is a Toeplitz matrix and
$I_{\mathrm{L}}$ is the identity matrix of dimension $\mathrm{L}$.
Obviously we also have
\begin{equation}
D_{\mathrm{L}}(\lambda)=\prod_{m=1}^{\mathrm{L}} (\lambda-\nu_m).
\label{exd}
\end{equation}
From the Cauchy residue theorem and analytical property of
\mbox{$e(x, \nu)$}, then $S(\rho_A)$ can be rewritten as
\begin{equation}
S(\rho_A)=\lim_{\epsilon \to 0^+}\frac{1}{2\pi \mathrm{i}}
\oint_{\Gamma'} \mathrm{d} \lambda\,  e(1+\epsilon, \lambda)
\frac{\mathrm{d}}{\mathrm{d} \lambda} \ln
D_{L}(\lambda)\;.\label{eaa}
\end{equation}
Here the contour \mbox{$\Gamma'$} in Fig~$\ref{fig1}$ encircles all
zeros of \mbox{$D_{\mathrm{L}}(\lambda)$} and function
\mbox{$e(1+\epsilon, \lambda)$} is analytic within the contour.
Just like Toeplitz matrix $\mathbf{G}_{\mathrm{L}}$ is generated by
function $\phi(\theta)$ in Eqs.~$\ref{mg1}$ and $\ref{mg2}$, Toeplitz
matrix $\widetilde{\mathbf{G}}_{\mathrm{L}}(\lambda)$ is generated
by function $\tilde{\phi}(\theta)$ defined by
\begin{equation}
\tilde{\phi}(\theta)=\left\{ \begin{array}{rl}
\lambda- 1, & -k_F < \theta <k_F,\\
\lambda+1, & k_F < \theta < (2\pi-k_F).
\end{array} \right.\label{xxsym}
\end{equation}
Notice that \mbox{$\tilde{\phi}(\theta)$} is a piecewise constant
function of $\theta$ on the unit circle, with jumps at
\mbox{$\theta=\pm k_F$}. Hence, if one can obtain the determinant of
this Toeplitz matrix analytically, one will be able to get a closed
analytical result for $S(\rho_A)$ which is our new result. Now, the
calculation of $S(\rho_A)$ reduces to the calculation of the
determinant of Toeplitz matrix
$\widetilde{\mathbf{G}}_{\mathrm{L}}(\lambda)$.

\subsection{$XY$ model}

Similarly let us introduce:
\begin{equation}
\widetilde{\mathbf{B}}_{L}(\lambda)=\mathrm{i}\lambda I_{L}-
\mathbf{B}_{L}, \quad D_{L}(\lambda)=\det
\widetilde{\mathbf{B}}_{L}(\lambda).
\end{equation}
 Here $I_{L}$ is the
identity matrix of dimension $2L$. By definition, we have
\begin{equation}
D_{L}(\lambda)=(-1)^{L} \prod_{m=1}^{L} (\lambda^2-\nu_m^2).
\label{exd1}
\end{equation}
Using again the Cauchy residue theorem  we obtain that, similar to (\ref{eaa}),
\begin{equation}
S(\rho_A)=\lim_{\epsilon \to 0^+} \frac{1}{4\pi \mathrm{i}}
\oint_{\Gamma'} \mathrm{d} \lambda\,  e(1+\epsilon, \lambda)
\frac{\mathrm{d}}{\mathrm{d} \lambda} \ln
D_{L}(\lambda)\;.\label{eaab}
\end{equation}
Here the contour \mbox{$\Gamma'$} in Fig~$\ref{fig1}$ encircles all
zeros of \mbox{$D_{L}(\lambda)$}.

\noindent We also realized that
$\widetilde{\mathbf{B}}_{L}(\lambda)$ is a block Toeplitz matrix
with the generator $\Phi(z)$, i.e.
\begin{equation}
\widetilde{\mathbf{B}}_L(\lambda)=\left( \begin{array}{cccc}
\widetilde{\Pi}_0 &\widetilde{\Pi}_{-1}& \ldots &\widetilde{\Pi}_{1-L}\\
\widetilde{\Pi}_{1}& \widetilde{\Pi}_0&   &   \vdots\\
\vdots &      & \ddots&\vdots\\
\widetilde{\Pi}_{L-1}& \ldots& \ldots& \widetilde{\Pi}_0
\end{array}     \right) \quad \textrm{with}\nonumber
\end{equation}
\begin{equation}
 \widetilde{\Pi}_l=\frac{1}{2\pi\mathrm{i}}\oint_{\Xi} \,
\mathrm{d} z\, z^{-l-1} \Phi(z), \quad \Phi(z)=\left(
\begin{array}{cc}
               \mathrm{i}\lambda & \phi(z)\\
               -\phi^{-1}(z)&\mathrm{i}\lambda
               \end{array} \right) \label{defphi}
\end{equation}
\begin{equation}
\textrm{and}\quad \phi(z)=
\left(\frac{\lambda_1^*}{\lambda_{1}}\frac{(1-\lambda_1\,
z)(1-\lambda_2\, z^{-1})}{(1-\lambda_1^* \, z^{-1})(1-\lambda_2^*\,
z)}\right)^{1/2}\label{defph}
\end{equation}

\begin{figure}[ht]
\begin{center}
\includegraphics[width=3in,clip]{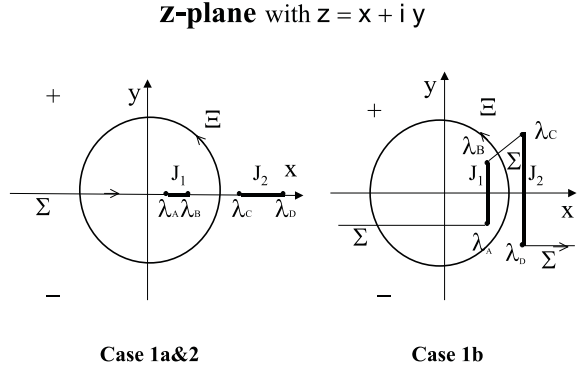}
\end{center}
\caption{\it Polygonal line $\Sigma$ (direction as labeled)
separates the complex $z$ plane into the two parts: the part
$\Omega_{+}$ which lies to the left of $\Sigma$, and the part
$\Omega_{-}$ which lies to the right of $\Sigma$. Curve $\Xi$ is the
unit circle in anti-clockwise direction. Cuts $J_1, J_2$ for
functions $\phi(z),w(z)$ are labeled by bold on line $\Sigma$.
Definition of the end points of the cuts $\lambda_{\ldots}$ depends
on the case: {\bf Case} $1$a:  $\lambda_A=\lambda_1$ and
$\lambda_B=\lambda_2^{-1}$, $\lambda_C= \lambda_2$ and $\lambda_D=
\lambda_1^{-1}$. {\bf Case} $1$b:
 $\lambda_A=\lambda_1$ and
$\lambda_B=\lambda_2^{-1}$, $\lambda_C= \lambda_1^{-1}$ and
$\lambda_D= \lambda_2$. {\bf Case } $2$:
 $\lambda_A=\lambda_1$ and $\lambda_B=\lambda_2$, $\lambda_C=
\lambda_2^{-1}$ and $\lambda_D= \lambda_1^{-1}$. $\P$} \label{fig2}
\end{figure}
We fix the branch by requiring that $ \phi(\infty)>0$. We use $*$ to
denote complex conjugation and $\Xi$ the unit circle shown in
Fig.~\ref{fig2}. $\lambda_1$ and $\lambda_2$ are defined differently
for different values of $\gamma$ and $h$. There are following  three
different cases:

\noindent   In {\bf Case} $1$a ($2\sqrt{1-\gamma^2}<h< 2$) and {\bf
Case} $2$ ($h> 2$), both $\lambda_1$ and $\lambda_2$ are real
\begin{equation}
\lambda_1=\frac{h-\sqrt{h^2-4(1-\gamma^2)}}{2(1+\gamma)},\quad
\lambda_2=\frac{1+\gamma}{1-\gamma} \lambda_1.\label{ldef1}
\end{equation}

\noindent  In {\bf Case} $1$b ($h^2<4(1-\gamma^2)$), both
$\lambda_1$ and $\lambda_2$ are complex
\begin{equation}
\lambda_1=\frac{h-\mathrm{i} \sqrt{4(1-\gamma^2)-h^2}}{2
(1+\gamma)},\quad
 \lambda_2=1/\lambda_1^*.\label{ldef2}
\end{equation}
Note that in the Case $1$ the poles of function $\phi(z)$
(Eq.~\ref{defph}) coincide with the points $\lambda_{A}$ and
$\lambda_{B}$, while in the Case $2$ they coincide with the points
$\lambda_{A}$ and $\lambda_{C}$.

\section{Block entropy of XX model and Fisher-Hartwig Formula}

From Eq. \ref{eaa}, one needs the calculation of Toeplitz
determinant$D_L(\lambda)$ with a singular generating function
\begin{equation}
\tilde{\phi}(\theta)=\left\{ \begin{array}{rl}
\lambda- 1, & -k_F < \theta <k_F,\\
\lambda+1, & k_F < \theta < (2\pi-k_F).
\end{array} \right.\label{xxsym1}
\end{equation}
It is easy to check that this function admits the canonical Fisher-Hartwig
factorization given by  Eq. \ref{fFH} with 
\begin{equation}\label{XXFH1}
m = 2, \quad \alpha_j = 0 \,\,\,\forall j, \quad \beta_0 = 0,
\quad \beta_2 = -\beta_1 \equiv \beta(\lambda) = 
\frac{1}{2\pi i}\ln\frac{\lambda+1}{\lambda-1},
\end{equation}
and
\begin{equation}\label{XXFH2}
e^{V(z)} \equiv e^{V_0} = (\lambda +1)\left(\frac{\lambda+1}{\lambda-1}\right)^{-k_{F}/\pi}.
\end{equation}
The branch of the logarithm is fixed by the condition, 
\begin{equation}
 -\pi \leq \arg \left(\frac{\lambda+1}{\lambda-1}\right) < \pi. \label{bbta3},
\end{equation}
For $\lambda \notin [-1,1]$, the left inequality is also strict, and hence  $|\Re (\beta_1(\lambda))
|<\frac{1}{2}$ and $| \Re (\beta_2(\lambda)) | < \frac{1}{2}$. Therefore,
Theorem 2 is applicable (indeed, even its earlier weaker version proven by
E. Basor \cite{B2}  would be suffice) and we see that 
the determinant $D_{\mathrm{L}}(\lambda)$ of
$\lambda I_{\mathrm{L}} -\mathbf{G}_{\mathrm{L}}$ can be
asymptotically represented as
\begin{eqnarray}
D_{\mathrm{L}}(\lambda)&=&\Bigl(2-2\cos(2
k_F)\Bigr)^{-\beta^2(\lambda)}
\left\{G\Bigl(1+\beta(\lambda)\Bigr) G\Bigl(1-\beta(\lambda)\Bigr)\right\}^2\nonumber \\
&&\left\{(\lambda+1)\Bigl((\lambda+1)/(\lambda-1)\Bigr)^{-k_F/\pi}\right\}^{\mathrm{L}}
\mathrm{L}^{-2 \beta^2(\lambda)}.\label{apd}
\end{eqnarray}
Here  $G$ is, as before,  the Barnes $G$-function and
\begin{equation}
G(1+\beta(\lambda))
G(1-\beta(\lambda))=e^{-(1+\gamma_E)\beta^2(\lambda)}\prod_{n=1}^{\infty}
 \left\{\left(1-\frac{\beta^2(\lambda)}{n^2}\right)^n e^{\beta^2(\lambda)/n^2} \right\}.
\end{equation}

Let us substitute the asymptotic form Eq. \ref{apd} into
Eq. \ref{eaa} and after some simplification\cite{jin},
 we have that
\begin{equation}
S(\rho_A)=\frac{1}{3} \ln \mathrm{L} +\frac{1}{6} \ln
\left(1-\left(\frac{h}{2}\right)^2\right) +\frac{\ln 2}{3}
+\Upsilon_1, ~~\mathrm{L}\to\infty \label{enl}
\end{equation}
with
\begin{eqnarray}
  \Upsilon_1& = &  - \int_0^\infty \mathrm{d} t \left\{ {e^{-t} \over 3 t}
  + {1 \over  t \sinh^2 (t/2)} - { \cosh (t/2) \over 2 \sinh^3 (t/2)}
  \right\}. \label{up12}
\end{eqnarray}
for $\mathrm{XX}$ model. The leading term of asymptotic of the
entropy $\frac{1}{3} \ln \mathrm{L}$ in Eq.~\ref{enl} was first
obtained based on numerical calculation and a simple conformal
argument in Ref.~\cite{vidal,LRV} in the context of entanglement. We
also want to mention that a complete conformal derivation for this
entropy was found in Ref.~\cite{K}. One can numerically evaluate
$\Upsilon_1$ to
 very high accuracy to be $0.4950179\cdots$.
 For zero magnetic field
 ($h=0$) case, the costant term $\Upsilon_1 +\ln 2/3$ for $S(\rho_A)$
is close to but different from $(\pi/3) \ln 2$, which can be found
by taking numerical accuracy to be more than five digits.

\section{Block entropy of $XY$ model and block Toeplitz determiniant}

For the block entropy of $XY$ model, by virtue of Eq.~\ref{eaab},
our objective becomes the asymptotic
 calculation of the determinant of block Toeplitz matrix
$ D_L(\lambda)$  or, rather, its $\lambda$ -derivative
$\frac{d}{d\lambda}\ln D_L(\lambda)$. 

Let us denote,
\begin{equation}\label{ztolambda}
z_1 := \lambda^{-1}_{1}, \quad \mbox{and}\quad z_2 := \lambda_2.
\end{equation}
It is easy  to check than that the generating function introduced in Eq. \ref{defphi}) -
Eq. \ref{defph} coincides with the one introduced  in Eq. \ref{defphi0}) -  Eq. \ref{defph0}
together with the case-separations and the $\lambda_A$ - $\lambda_D$ labeling
of the branch points. Hence one can use Theorem 5 and substitute the asymptotic form
Eq. \ref{block3} into Eq. \ref{eaab}. Deforming the original contour of integration
to the contour $\Gamma$ as indicated  in Fig.~\ref{fig1} we arrive
at the following expression for the {\it entropy} \cite{ijk1,ijk2}:
\begin{equation}\label{33} S(\rho_{A})= \frac{1}{2}\int_{1}^{\infty}\ln
\left(\frac{\theta_{3}\left(\beta(\lambda) + \frac{\sigma
\tau}{2}\right) \theta_{3}\left(\beta(\lambda) - \frac{\sigma
\tau}{2}\right)} {\theta^{2}_{3}\left(\frac{\sigma
\tau}{2}\right)}\right)\, d\lambda,
\end{equation}
wich can also be written in the form,
\begin{equation}\label{333}
S(\rho_{A})= \frac{\pi}{2} \int_{0}^{\infty}\ln
\left(\frac{\theta_{3}\left(\mathrm{i} x + \frac{\sigma
\tau}{2}\right) \theta_{3}\left(\mathrm{i} x -\frac{\sigma
\tau}{2}\right)} {\theta^2_{3}\left(\frac{\sigma
\tau}{2}\right)}\right) \frac{dx}{\sinh^2(\pi x)}
\end{equation}
This is a limiting expression as $L \to \infty$. In \cite{ijk2} it is also proven  that
 the corrections in
Eq.~\ref{33}  are of order of $
O\left({\lambda_{C}^{-L}}/{\sqrt{L}}\right).$ 

The entropy has singularities at {\it phase transitions}. When $\tau
\to 0$ we can use   Landen transform (see  \cite{ww}) to get the
following estimate of the theta-function for small $\tau$ and pure
imaginary $s$:
$$\ln \frac{\theta_{3}\left(s \pm
\frac{\sigma \tau}{2}\right)}{\theta_{3}\left(\frac{\sigma
\tau}{2}\right)} = \frac{\pi}{i\tau}s^{2} \mp \pi i \sigma s +
O\left(\frac{e^{-i\pi/\tau}}{\tau^{2}} s^2\right), ~\textrm{as $\tau
\to 0$}.$$ Now the leading term in the expression for  the entropy
(\ref{33})
 can be replaced   by
\begin{equation}\label{4} S(\rho_{A}) = \frac{\mathrm{i}\pi}{6\tau}+
 O\left(\frac{e^{-i\pi/\tau}}{\tau^{2}}\right)
\quad \textrm{for $\tau \to 0$}.
\end{equation}
Let us consider two physical situations corresponding to small
$\tau$ depending on the case defined on the page 2:

\begin{enumerate}
\item{\it Critical magnetic field}:  $\gamma\neq 0$ and $h\to 2$.

This  is included  in our   Case
 $1$a  and  Case $2$, when $h> 2\sqrt{1-\gamma^2 }  $.
As $h\to 2$ the end points of the cuts  $\lambda_B \to \lambda_C$,
so $\tau$ given by Eq.~(\ref{module}) simplifies and
 we obtain from Eq.~(\ref{4}) that the entropy is:
\begin{equation} S(\rho_{A}) = -\frac{1}{6} \ln |2-h| + \frac{1}{3}\ln
4\gamma, \label{cardy} \quad \textrm{for $h\to 2$ and $\gamma\neq
0$}
\end{equation}
correction is $O(|2-h|\ln^{2}|2-h|) $. This limit agrees with
predictions of conformal approach \cite{K,cardy}.
 The first term in the right hand side of (\ref{cardy})
can be represented as $(1/6)\ln \xi$, this confirms a conjecture of
 \cite{cardy}.  The correlation length $\xi$
  was  evaluated in \cite{mccoy}.

\item{\it An approach to $XX$ model}:  $\gamma\to 0$ and $h<2$: It
is included in Case $1$b, when  $0<h<2\sqrt{1-\gamma^2} $. Now
 $\lambda_B \to \lambda_C$ and  $\lambda_A \to \lambda_D$, we can calculate $\tau$ explicitly. The entropy becomes:
\begin{equation}
S^{0}(\rho_{A}) = -\frac{1}{3} \ln \gamma + \frac{1}{6}{\ln
(4-h^2)}+\frac{1}{3}\ln 2, \quad\textrm{for $\gamma\to 0$ and $h<2$}
\end{equation}
correction is $O(\gamma\ln^2 \gamma) $. This agrees with  \cite{jin} (see also Eq.~\ref{enl}).
\end{enumerate}

As it has already been indicated,  the theta-functions involved in the asymptotic formula Eq. (\ref{block3})
has zeros at the points $\pm \lambda_m$ which are defined in Eq.~(\ref{zeros}). Theorem 5
shows, in particular, that  in the large $L$ limit,
the points $\pm\lambda_{m}$  are double zeros  of the
$D_L  (\lambda) $. More precisely, we see that in the
 large $L$ limit the eigenvalues $\nu_{2m}$ and  $\nu_{2m+1}$ from
  (\ref{eaap})  merge to $\lambda_{m}$:
\begin{equation}\label{doubleMay}
 \nu_{2m}, \nu_{2m+1} \to \lambda_{m},
\end{equation}
which in turn implies (cf. Eq. \ref{eaap})  the following equivalent description of the limiting
entropy $S(\rho_{A})$ \cite{ijk1}:

{\it The limiting entropy, $S(\rho_{A})$, of the subsystem
can be identified with the infinite
convergent series,
\begin{equation}\label{3333May}
 \diamondsuit  \quad \qquad  S(\rho_A) = \sum_{m=-\infty}^{\infty} e(1, \lambda_m) =\sum_{m=-\infty}^{\infty}
  (1+\lambda_{m})\ln \frac{2}{1+\lambda_{m}}  \qquad  \diamondsuit
  \end{equation}}
  
\noindent
Indeed, equation (\ref{3333May}) follows from the substitution of  Eq. (\ref{block3}) into Eq. (\ref{eaab})
and integrating over the {\it original} contour $\Gamma'$ of
 Fig.~\ref{fig1}

It is also worth mentioning that relation (\ref{doubleMay}) also
indicates the degeneracy of the spectrum of the matrix ${\bf B}_{L}$
and  an appearance of an {\it extra symmetry} in the large $L$
limit.

\noindent
{\bf Remark}. These numbers $\lambda_{m}$  satisfy an estimate:
$$
|\lambda_{m+1} - \lambda_{m}| \leq 4\pi \tau_{0} \quad \mbox{with}
\quad \tau_{0} = -i\tau.
$$
This means that  $(\lambda_{m+1} - \lambda_{m}) \to 0$ as $\tau \to
0$ for every $m$. This is useful for understanding of  large $L$
limit of the $XX$ case corresponding  to $\gamma \to 0$, as
considered in \cite{jin}.  The estimate explains why in the $XX$
case  the singularities of the logarithmic derivative of the
Toeplitz determinant
 $d\ln D_L(\lambda)/ d\lambda $ form
a cut along the interval $[-1, 1]$, while in the $XY$ case it has  a
discrete set of poles at points $\pm \lambda_{m}$ of
Eq.~(\ref{zeros}).

The higher genus analog of formula Eq.~(\ref{33}) for the class
of  quantum spin chains introduced  by J.\  Keating and
F.\ Mezzadri in  \cite{keat} has been obtained in \cite{imm}.

\noindent
{\bf Remark.} 
It was shown by Peschel in \cite{pes} (who also suggested an
alternative heuristic derivation of equation (\ref{3333May}) based
on the work  \cite{cardy}), the series
(\ref{3333May}) can be summed up to an elementary function of the
complete elliptic integrals corresponding to the modular parameter
$\tau$ - see Eqs \ref{pes1} and \ref{pes2} below. It is an open problem whether an analogous representation of the
integral Eq.~(\ref{33}) exists for higher genus. The key issue here is the 
extreme complexity of the identification of the zero divisor of the theta-
functions in the dimension grater than 1.

\section{Renyi entropy and the spectrum of reduced density matrix of XY model}

The Renyi entropy of $S_{\alpha}(\rho_A)$ of the block of spins is defined by the
expression
\begin{equation}
S_{\alpha}(\rho_A)= \frac{1}{1-\alpha}\ln Tr(\rho_A^{\alpha}),
\qquad \alpha\neq 1 ~~\textrm{and}~~ \alpha>0 .\label{olds1}
\end{equation}
Here the power $\alpha$ is a parameter. The Renyi entropy is intimately related to
the spectrum of the reduced density matrix $\rho_A$. Indeed, let $\lambda_n$,
($0<\lambda_n<1$) and  $a_n$  denote the eigenvalues and their multiplicities
of the operator $\rho_A$. The spectrum is completely determined by its momentum function, i.e.
by the $\zeta$-function of $\rho_A$ ,
\be\label{zeta}
\zeta_{\rho_A}(\alpha) = \sum_{n=0}^{\infty} a_n\lambda_n^{\alpha} .
\ee
The obvious equation takes place,
\be\label{mS}
\zeta_{\rho_A}(\alpha) = e^{(1-\alpha)S_R(\rho_{A},\alpha)}.
\ee
The key point is that we can evaluate $S_{\alpha}(\rho_A)$, and hence  $\zeta_{\rho_A}(\alpha)$,
explicitly. 

As it is shown in \cite{jin},  the Renyi entropy   $S_{\alpha}(\rho_A)$ of a block of $L$ neighboring spins,
before the large $L$ limit is taken,  can be represented by the finite sum,
\be \label{renyiLdef}
 S_R (\rho_A,\alpha) = {1 \over 1- \alpha} \sum_{k=1}^L
   \ln \left[ \left( {1 + \nu_k \over 2 } \right)^\alpha
   + \left( {1 - \nu_k \over 2 } \right)^\alpha \right] ,
\ee
where the numbers
$$
\pm i\nu_{k}, \quad k = 1, ..., L
$$
are the eigenvalues of the same  block Toeplitz matrix Eq.~\ref{exd1}
as we worked with in Section 3.4. In virtue of Eq.~(\ref{doubleMay}), 
 the Renyi entropy in the large $L$ limit can be identified
with the convergent series,
\be \label{rendef0}
   S_R (\rho_A,\alpha) = {1 \over 1-\alpha} \sum_{m=-\infty}^\infty
   \ln \left[ \left( {1 + \lambda_m \over 2 } \right)^\alpha
   + \left( {1 - \lambda_m \over 2 } \right)^\alpha \right] ,
\ee with
\begin{equation}
  \lambda_{m} =
  \tanh \left(m + \frac{1-\sigma}{2}\right)\pi \tau_{0}. \label{lam}
\end{equation}
The summation of the series can be done following the same approach
as in \cite{pes} in the case of the von Neuman entropy. The result is (for details
see \cite{fik}) the following,
\begin{equation}\label{renyi1}
S_R(\rho_{A},\alpha)
= \frac{\alpha}{1-\alpha}\left(\frac{\pi \tau_0}{12} +\frac{1}{6}\ln\frac{k\;k'}{4}\right)
+\frac{1}{1-\alpha}\ln \prod_{n=0}^{\infty}\left(1 + q_{\alpha}^{2n+1}\right)^2,
\ee
\be\label{qalpha}
q_{\alpha} = e ^{-\alpha\pi\tau_0},
\ee
for the case $h>2$, and
$$
S_R(\rho_{A},\alpha)
= \frac{\alpha}{1-\alpha}\left(-\frac{\pi \tau_0}{6} +\frac{1}{6}\ln\frac{k'}{4k^2}\right)
+\frac{1}{1-\alpha}\ln \prod_{n=1}^{\infty}\left(1 + q_{\alpha}^{2n}\right)^2
$$
\be\label{renyi2}
+\frac{1}{1-\alpha}\ln 2,
\ee
$$
q_{\alpha} = e ^{-\alpha\pi\tau_0},
$$
for the case $h<2$. In these equations, 
$\tau_0 \equiv -i\tau$ is  the module parameter  defined in Eq.~(\ref{module},) and $k \equiv k(q_1)$,
$k' \equiv k'(q_1)$ are the standard elliptic modular functions, see e.g. \cite{ww}. 
The quantities  $k$ and $k'$ are simply related to the basic physical parameters
$\gamma$ and $h$. Indeed, one has that
\begin{eqnarray}
  \label{kmain}
  k \equiv \left\{ \begin{array}{ll}
  \sqrt{(h/2)^2+\gamma^2-1}\; /\; \gamma \; ,&
  \mbox{Case 1a:~$4(1-\gamma^2)<h^2<4$;} \\
  \sqrt{({1-h^2/4-\gamma^2})/({1-h^2/4})} \; , &
  \mbox{Case 1b:~$h^2<4(1-\gamma^2)$;} \\
  \gamma\; / \;\sqrt{(h/2)^2+\gamma^2-1} \; , &
  \mbox{Case 2~:~$h>2$.}
  \end{array} \right.,
\end{eqnarray}
$$
k' = \sqrt{1 - k^2}.
$$
By standard techniques of the theory of elliptic functions, equation Eq.~(\ref{module}) can be
transformed into the following representation for  the module $\tau_0$ as a function of $k$.
\be
   \label{taudef}
   \tau_0 \equiv \frac{I(k')}{I(k)} \; ,
   \qquad \qquad k'=\sqrt{1-k^2},
\ee
$I(k)$ is the complete elliptic integral of the first kind,
\begin{equation}\label{ellint}
I(k) = \int_{0}^{1}\frac{dx}{\sqrt{(1-x^2)(1 - k^{2}x^{2})}}.
\end{equation}

The $q$-products in Eq. \ref{renyi1} and Eq. \ref{renyi2} can be expressed in terms of 
 the {\it elliptic lambda function } or
{\it $\lambda$ - modular function}.  The $\lambda$ - function  plays a central
role in the theory of modular functions and modular forms, and it is  defined by the equation
(see  e.g. \cite{ww}), 
\be \label{kappadef}
\lambda(\tau) =  \frac{\theta^{4}_{2}(0|\tau)}{\theta^{4}_{3}(0|\tau)}
\equiv k^2(e^{i\pi \tau}), \quad \Im \tau > 0,
\ee
where $\theta_{j}(s|\tau)$, $j = 3, 4$ are Jacobi theta-fucntions;
the function $\theta_{3}(s|\tau)$ has already been defined
in Eq. \ref{jac}, while the function $\theta_{4}(s|\tau)$ is defined by the equation,
\begin{equation}\label{theta4}
\theta_(s|\tau)=\sum_{n=-\infty}^{\infty}(-1)^{n} e^{\pi i \tau n^2+2\pi i s
n}.\label{jac4}
\ee
The $\lambda$ - function is analytic function of $\tau$,  $\Im \tau >0$,
and it satisfies the following symmetry
relations with respect to the actions of the generators of the
modular group,
\be \label{kappamod1}
\lambda(\tau + 1) = \frac{\lambda(\tau)}{\lambda(\tau) - 1},
\ee
\be \label{kappamod2}
\lambda\left(-\frac{1}{\tau}\right) = 1 - \lambda(\tau).
\ee

In terms of the $\lambda$ - modular  function, the formulae
for Renyi read as follows \cite{fik}.
\begin{eqnarray}
	&&S_R(\rho_{A},\alpha) \label{renlambda1}\\
	   &=& {1 \over 6} \; { \alpha \over 1- \alpha } \; \ln \left( k \; k' \right)
	   - {1 \over 12} \; { 1 \over 1-\alpha} \; \ln \Bigl(\lambda(\alpha i \tau_{0})
	   (1-\lambda(\alpha i \tau_{0}))\Bigr)
	    + {1 \over 3} \ln 2, \nonumber
\end{eqnarray}
for $h> 2$ and
\begin{eqnarray}
   &&S_R(\rho_{A},\alpha) \label{renlambda2}\\
   &=& {1 \over 6} \; {\alpha \over 1-\alpha } \; \ln
   \left( {k'\over k^2 } \right) + {1 \over 12} \; {1 \over 1-\alpha } \;
   \ln \frac { \lambda^2(\alpha i \tau_{0})}{1-\lambda(\alpha i \tau_{0})}
   + {1 \over 3} \ln 2, \nonumber
\end{eqnarray}
for $h < 2$.

Eqs. \ref{renlambda1} and  \ref{renlambda2} allow to apply to the study of the Renyi
entropy the apparatus of the theory of modular functions.

\noindent
{\bf Remark.} Using Eq. \ref{renlambda1} and Eq. \ref{renlambda2} one can evaluate the asymptotics
of the Renyi entropy as $\alpha \to 1$. This would lead to the following formulae for the 
Neumann entropy,
\begin{eqnarray}
   S(\rho_A)=   \frac {1} {6} \left [\;\ln{ (\frac {k^2} {16 k'})} + (1-\frac {k^2} {2})
           \frac {4 I(k) I(k')} {\pi} \right ] + \ln\;2 , \label{pes1}
   \end{eqnarray}
  in {\bf Case 1}, and
  \begin{eqnarray}
     S(\rho_A)=  \frac {1} {12} \left [\;\ln{ (\frac {16} {(k^2 k'^2)}} + (k^2-k'^2)
           \frac {4 I(k) I(k')} {\pi} \right ],\label{pes2}
     \end{eqnarray}
  in {\bf Case 2}. For the Cases 1a and 2 these formulae were first obtained by Peschel 
  in \cite{pes} by a direct summation of series (\ref{3333May})

\section{Spectrum of the limiting density matrix}
We will show now, following \cite{fikt}, how to extract from  Eq.~(\ref{renyi1}) and (\ref{renyi2})
the information about the spectrum of the density matrix $\rho_A$.

Consider first the case $h>2$.  Combining equations (\ref{renyi1}) and  (\ref{mS}), we arrive at the following
representation for the $\zeta$-function $\zeta_{\rho_A}(\alpha)$,
\be\label{zeta1}
\zeta_{\rho_A}(\alpha) = e^{\alpha\left(\frac{\pi \tau_0}{12} +\frac{1}{6}\ln\frac{k\;k'}{4}\right)}
\prod_{n=0}^{\infty}\left(1 + q_{\alpha}^{2n+1}\right)^2.
\ee
At the same time,  using the classical arguments of the theory of partitions
(see e.g. \cite{andrews}, Chapter 11, equation (11.1.4)) we have that
\be\label{fourier1}
\prod_{n=0}^{\infty}\left(1 + q^{2n+1}\right) = \sum_{n=1}^{\infty}p_{\cal O}^{(1)}(n)q^n,
\ee
where $p_{\cal O}^{(1)}(0) = 1$ and   $p_{\cal O}^{(1)}(n)$, for $n >1$, denote the number of
partitions of  $n$ into distinct positive odd integers, i.e.
\begin{eqnarray}
	\#\left\{(m_1, ..., m_k): m_j = 2r_j + 1, \quad m_1 > m_2 > ... > m_k, \right.\nonumber\\ \left.
	\quad n = m_1 + m_2 +  ... + m_k\right\}.\nonumber
\end{eqnarray}
Hence (\ref{zeta1}) becomes,
\be\label{zeta2}
\zeta_{\rho_A}(\alpha) = e^{\alpha\left(\frac{\pi \tau_0}{12} +\frac{1}{6}\ln\frac{k\;k'}{4}\right)}
\sum_{n=0}^{\infty}a_n q_{\alpha}^{n},
\ee
where,
\be\label{adef}
a_0 = 1, \quad a_n = \sum_{l=0}^{n}p_{\cal O}^{(1)}(l)p_{\cal O}^{(1)}(n-l).
\ee
Finally, observing that
\be\label{lambdan}
q_{\alpha}^n = \left(e^{-\pi\tau_0n}\right)^{\alpha},
\ee
we conclude that
\be\label{zeta3}
\zeta_{\rho_A}(\alpha)  = \sum_{n=0}^{\infty}a_n \lambda_n^{\alpha}, \quad \lambda_n = 
e^{-\pi\tau_0n +\frac{\pi \tau_0}{12} + \frac{1}{6}\ln \frac{k\;k'}{4}}.
\ee
Comparing the last equation with equation (\ref{zeta}) we arrive at the
following theorem.

\noindent
{\bf Theorem 6.} (\cite{fikt}) {\it Let the magnetic field $h >2$. Then,  the eigenvalues  of the reduced density matrix 
$\rho_{A}$ are given by the equation,
\be\label{spectrumfinal}
\lambda_n = e^{-\pi\tau_0n +\frac{\pi \tau_0}{12} + \frac{1}{6}\ln \frac{k\;k'}{4}}, \quad n = 0, 1, 2, ....,
\ee
and the corresponding multiplicities $a_n$ are determined by the relation (\ref{adef}).}

The case $h < 2$ is treated in a very similar way. Instead of (\ref{zeta1}) we have now 
the formula,
\be\label{zeta4}
\zeta_{\rho_A}(\alpha) = 2e^{\alpha\left(-\frac{\pi \tau_0}{6} +\frac{1}{6}\ln\frac{k'}{4k^2}\right)}
\prod_{n=0}^{\infty}\left(1 + q_{\alpha}^{2n}\right)^2,
\ee
where $q_{\alpha}$ as in (\ref{qalpha}). The analog of the Taylor expansion (\ref{fourier1}) 
is the equation,
\be\label{fourier2}
\prod_{n=0}^{\infty}\left(1 + q^{2n}\right) = \sum_{n=1}^{\infty}p_{\cal N}^{(1)}(n)q^{2n},
\ee
where $p_{\cal N}^{(1)}(0) = 1$ and   $p_{\cal N}^{(1)}(n)$, for $n >1$, denote the number of
partitions of  $n$ into distinct positive  integers, i.e.
$$ 
 \#\left\{(m_1, ..., m_k):  m_1 > m_2 > ... > m_k \geq 0,
\quad n = m_1 + m_2 +  ... + m_k\right\}.
$$
Hence (\ref{zeta4}) becomes,
\be\label{zeta5}
\zeta_{\rho_A}(\alpha) = 2e^{\alpha\left(-\frac{\pi \tau_0}{6} +\frac{1}{6}\ln\frac{k'}{4k^2}\right)}
\sum_{n=0}^{\infty}b_n q_{\alpha}^{2n},
\ee
where,
\be\label{bdef}
b_0 = 1, \quad b_n = \sum_{l=0}^{n}p_{\cal N}^{(1)}(l)p_{\cal N}^{(1)}(n-l).
\ee
Finally, observing that
\be\label{lambda2n}
q_{\alpha}^{2n} = \left(e^{-2\pi\tau_0n}\right)^{\alpha},
\ee
we conclude that
\be\label{zeta6}
\zeta_{\rho_A}(\alpha)  = 2\sum_{n=0}^{\infty}b_n \lambda_n^{\alpha}, \quad \lambda_n = 
e^{-2\pi\tau_0n -\frac{\pi \tau_0}{6} +\frac{1}{6}\ln\frac{k'}{4k^2}}.
\ee
Comparing the last equation again with Eq. \ref{zeta} we arrive at the
analog of Theorem 6 for the case $h<2$ .

\noindent
{\bf Theorem 7.} (\cite{fikt}) {\it Let the magnetic field $h < 2$. Then,  the eigenvalues  of the reduced density matrix 
$\rho_{A}$ are given by the equation,
\be\label{spectrumfinal2}
\lambda_n = e^{-2\pi\tau_0n -\frac{\pi \tau_0}{6} +\frac{1}{6}\ln\frac{k'}{4k^2}},
\quad n = 0, 1, 2, ....,
\ee
and the corresponding multiplicities equal $2b_n$ where the integers $b_n$ are determined by the relation (\ref{bdef}).}

Let
\begin{equation}\label{genfunk}
f(x) := \sum_{n=0}^{\infty}a_nx^n,
\ee
be the generating function for the coefficients $a_n$. Then, Eq. \ref{renlambda1} and
Eq. \ref{mS} in conjunction with  the symmetry property (\ref{kappamod2})
allow to analyze the asymptotic behavior of the functio $f(x)$ generating function
as $x \to 1$. In its turn, this fact yields the evaluation of the large $n$ asymptotics of the 
multiplicities $a_n$ (details are in  \cite{fikt}).

\noindent
{\bf Theorem 8.} {\it Let $a_n$ be the multiplicities of the eigenvalues of the
reduced density matrix for $h >2$. Then their large $n$ behavior is given
by the relation,
\be\label{asymp12}
a_n\sim 2^{-3/2}3^{-1/4}n^{-3/4}e^{\pi \sqrt{\frac{n}{3}}}, \quad n\to \infty.
\ee}
 \section{Summary and Open Problems}
We want to emphasise that the method described here  also works for evaluation of correlation functions. For example space, time and temperature dependent correlation function of quantum spins was evaluated in \cite{sla}. The book \cite{korepin} explains how to apply this method for calculation of correlation functions in Bose gas with delta interaction. 

On the other hand there are still {\bf open problems}. 
For example let us consider the $XXZ$ model. The Hamiltonian can be written in terms of Pauli matrices $\sigma_{n}$:
\begin{eqnarray}
	H_{XXZ}=-\sum^{\infty}_{n=-\infty}\sigma^{x}_{n}\sigma^{x}_{n+1}+\sigma^{y}_{n}\sigma^{y}_{n+1}+\Delta\sigma^{z}_{n}\sigma^{z}_{n+1}. \label{XXZ}
\end{eqnarray}
At $\Delta<-1$ the model has a gap and the ground state is anti-ferromagnetic. 
Challenging problem is  to  calculate the von Neumann entropy and R\'enyi entropy of large block of spins on the infinite lattice. It will be interesting to find the dependence of limiting entropy on $\Delta$.

\begin{acknowledgements}
We  thank S.\ Bravyi,   I. Krasovsky and   B.M.\ McCoy  for useful  discussions. We appreciate collaboration with   F.\ Franchini and  L.A. Takhtajan. 
This work was supported by National Science Foundation (USA) under grants DMS-0503712, PHY-9988566  and  DMS-0701768 and by the EPSRC (UK) grant 
No. EP/F014198.
\end{acknowledgements}



\end{document}